\documentclass[letterpaper,aps,prl,superscriptaddress,showpacs,floatfix,twocolumn]{revtex4}

\usepackage{graphicx}
\usepackage{hyperref}
\usepackage{amsmath}
\usepackage{url}
\usepackage{colordvi}
\usepackage{subfigure}
\usepackage{times}

\begin{document}

\title{Lepton angular distribution of $Z$ boson production and 
jet discrimination}

\author{Jen-Chieh Peng}
\affiliation{Department of Physics, University of Illinois at
Urbana-Champaign, Urbana, Illinois 61801, USA}

\author{Wen-Chen Chang}
\affiliation{Institute of Physics, Academia Sinica, Taipei 11529, Taiwan}

\author{Randall Evan McClellan}
\affiliation{Department of Physics, University of Illinois at
Urbana-Champaign, Urbana, Illinois 61801, USA}

\affiliation{Thomas Jefferson National Accelerator Facility,
Newport News, Virginia 23606, USA}

\author{Oleg Teryaev}
\affiliation{Bogoliubov Laboratory of Theoretical Physics,
JINR, 141980 Dubna, Russia}

\date{\today}

\begin{abstract}
High precision data of lepton angular distributions in inclusive $Z$ 
boson production, reported
by the CMS and ATLAS Collaborations, showed pronounced 
transverse momentum ($q_T$) dependencies 
of the $A_0$ and $A_2$ coefficients. 
Violation of the 
Lam-Tung relation, $A_0 = A_2$, was also found. An intuitive understanding
of these results can be obtained from a geometric approach.
We predict that $A_0$ and $A_2$ for $Z$ plus single gluon-jet events 
are very different from that of $Z$ plus single quark-jet events, allowing
a new experimental tool for checking various
algorithms which attempt to discriminate quark jets from gluon jets.
We also predict that the Lam-Tung relation would
be more severely violated for the $Z$ plus multiple-jet data than
what has been observed so far for inclusive $Z$ production data.
These predictions can be readily tested using existing LHC data.
\end{abstract}
\keywords{Z-boson production,Z decay angular distribution,jet discrimination}
\pacs{12.38.Lg,14.20.Dh,14.65.Bt,13.60.Hb}

\maketitle

Measurement of lepton angular distribution in $W$ and $Z$ boson
production has long been advocated as a sensitive tool for
understanding the production mechanism of these gauge
bosons~\cite{mirkes94,berger98}. The lepton angular
distribution in $Z$ boson production was first measured
by the CDF Collaboration
for $\bar p p$ collision at 1.8 TeV~\cite{cdf}.
More recently, the CMS~\cite{cms} and
ATLAS~\cite{atlas} Collaborations at LHC
reported high-statistics measurements of the
lepton angular distribution of $Z$ boson production in $p p$ collision
at $\sqrt s = 8$ TeV. Pronounced $q_T$ dependencies, where $q_T$ refers
to the transverse momentum of $Z$ boson, were observed
for the lepton angular distributions. The Lam-Tung relation~\cite{lam78},
which is the analog of the Callan-Gross relation~\cite{callan69}
in deep-inelastic
scattering, was found to be significantly violated~\cite{cms,atlas}.

In a recent analysis~\cite{peng16,chang17} of the LHC $Z$ boson
angular distribution data,
we showed that the $q_T$
dependence of lepton angular distributions can be well described
by an intuitive
geometric approach.
These data were shown to be sensitive to the relative contributions
between the $q \bar q$ annihilation and the $qg$ Compton process.
The violation of the Lam-Tung relation was attributed~\cite{peng16}
to the acoplanarity between the `hadron plane' and the `quark plane',
to be defined later.
The magnitude of the violation of the Lam-Tung relation
was shown to depend on the amount of the acoplanarity.

The angular distribution data presented by the CMS and ATLAS
Collaborations correspond to inclusive $Z$ boson production.
For $Z$ boson produced with a sizable $q_T$ there must be accompanying
single jet or multiple jets to balance the $q_T$
of the $Z$-boson. In this paper we show that new insight on
the $q_T$ dependence of the angular distribution coefficients, as well
as the violation of the Lam-Tung violation, could be
obtained if the angular distribution coefficients
were analyzed as a function of the number of accompanying jets.
We also show that the angular distribution coefficients
for $Z$ plus single jet data would provide a powerful tool
for testing various algorithms
designed to distinguish quark jets from gluon jets.

\begin{figure}[tb]
\includegraphics*[width=\linewidth]{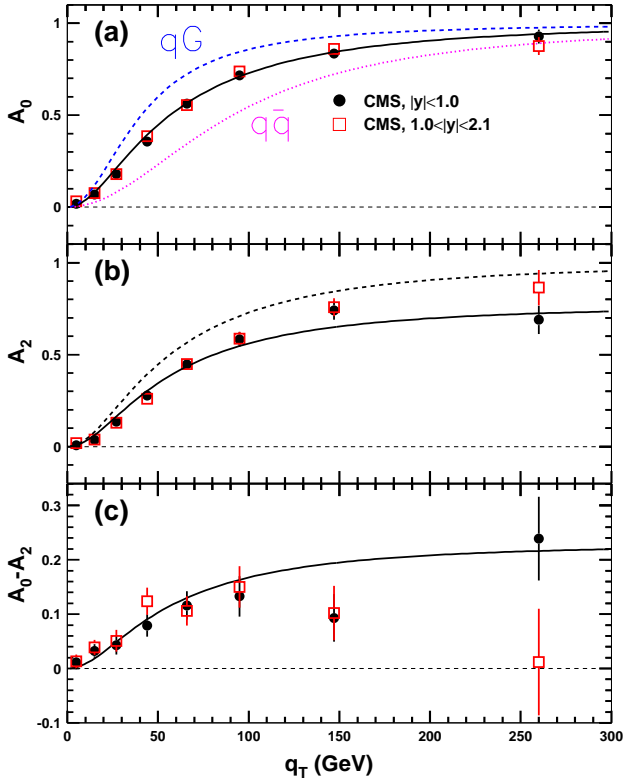}
\caption{The CMS data on $A_0$, $A_2$ and $A_0 - A_2$
measured at two rapidity ($y$) regions. The solid curves correspond to
calculations based on the geometric model discussed in the text. The dotted and
dashed curves in (a) are calculations for the $q \bar q$ and $qg$ processes,
respectively. The dashed curve in (b) corresponds to the Lam-Tung relation,
$A_0 = A_2$, where $A_0$ is taken from the solid curve in (a).}
\label{fig1}
\end{figure}

The lepton angular distribution in the $Z$ rest frame can be
expressed as~\cite{cms,atlas}
\begin{eqnarray}
\frac{d\sigma}{d\Omega} & \propto & (1+\cos^2\theta)+\frac{A_0}{2}
(1-3\cos^2\theta)+A_1 \sin 2 \theta\cos\phi \nonumber \\
& + & \frac{A_2}{2} \sin^2\theta \cos 2 \phi
+ A_3 \sin\theta \cos\phi + A_4 \cos\theta \nonumber \\
& + & A_5 \sin^2\theta \sin 2\phi
+ A_6 \sin 2\theta \sin\phi \nonumber \\
& + & A_7 \sin\theta \sin\phi,
\label{eq:eq1}
\end{eqnarray}
where $\theta$ and $\phi$ are the polar and azimuthal angles of $l^-$
($e^-$ or $\mu^-$) in the rest frame of $Z$. The original
Drell-Yan model~\cite{drell} neglected QCD effects and intrinsic transverse momenta of the
annihilating quark and antiquark. Hence, the angular
distribution is simply $1+\cos^2\theta$ and all angular distribution
coefficients, $A_i$, vanish. For non-zero dilepton transverse momentum,
$q_T$, these coefficients can deviate from zero. However, it was
predicted that the coefficients $A_0$ and $A_2$ should remain
identical, $A_0 = A_2$, which is the Lam-Tung relation~\cite{lam78}.
The high-statistics $Z$ boson production
data from the LHC allow a precise test of the Lam-Tung relation.
Figure~\ref{fig1} shows the CMS data for $A_0$, $A_2$, and
$A_0 - A_2$ measured at two rapidity ($y$) regions. Pronounced $q_T$
dependence of
$A_0$ and $A_2$ is observed. Moreover, the Lam-Tung relation,
$A_0 - A_2=0$, is found to be clearly violated.

\begin{figure}[tb]
\includegraphics*[width=\linewidth]{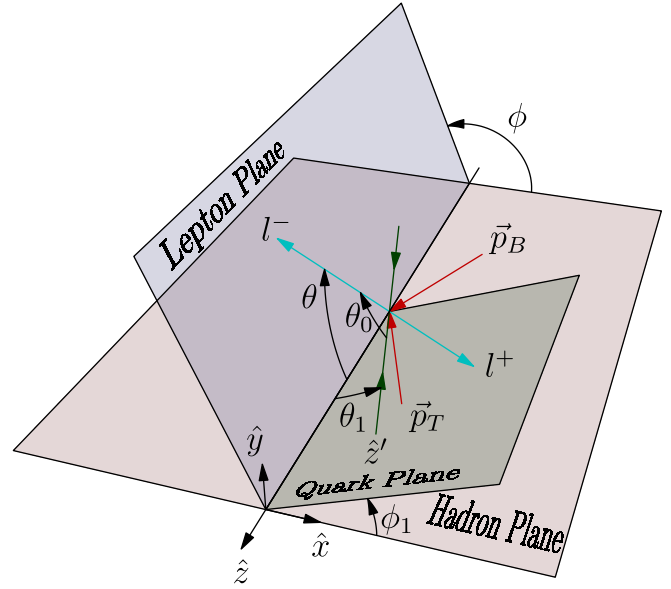}
\caption{Definition of the Collins-Soper (C-S) frame and various angles
and planes in the rest frame of $Z$ boson. The hadron plane
is formed by $\vec P_B$ and $\vec P_T$, the momentum vectors of
the colliding hadrons $B$ and $T$. The $\hat x$ and $\hat z$ axes
of the C-S frame both lie in the hadron plane with
$\hat z$ axis bisecting the $\vec P_B$ and $- \vec P_T$ vectors.
The quark ($q$) and antiquark ($\bar q$) annihilate collinearly
with equal momenta to form the $Z$ boson, while the quark momentum
vector $\hat z^\prime$ and the $\hat z$ axis form the quark plane.
The polar and azimuthal angles of $\hat z^\prime$ in the Collins-Soper
frame are $\theta_1$ and $\phi_1$. The $l^-$ and $l^+$ are emitted
back-to-back with $\theta$ and $\phi$ specifying the polar and
azimuthal angles of $l^-$.}
\label{fig2}
\end{figure}

To provide some insight on the meaning of various
angular distribution coefficients $A_i$ in
Eq.~(\ref{eq:eq1}),
we first present a derivation for Eq.~(\ref{eq:eq1}) based on an intuitive
geometric picture~\cite{peng16,chang17}. In the frame where $Z$ is at rest,
we define three different planes, namely, the
hadron plane, the quark plane, and the lepton plane,
shown in Fig.~\ref{fig2}.
For non-zero $q_T$, the momenta of the colliding hadrons,
$\vec P_B$ and $\vec P_T$, are
no longer collinear and they
form the ``hadron plane" shown in Fig.~\ref{fig2}.
Various coordinate systems have been considered in the literature, and
the Collins-Soper (C-S) frame~\cite{cs} was used by both the
CMS and ATLAS
Collaborations. For the C-S
frame, both the $\hat x$ and $\hat z$ axes lie in the hadron plane,
and the
$\hat z$ axis bisects $\vec P_B$ and $- \vec P_T$ with an angle $\beta$.
It is straightforward to show that
\begin{equation}
\tan \beta = q_T / Q,
\label{eq:eq2}
\end{equation}
where $Q$ is the mass of the $Z$ boson. Equation (2) shows that $\beta$
vanishes at $q_T = 0$, as $\vec P_B$ and $\vec P_T$ are collinear at
this limit. For
non-zero $q_T$, $\beta$ increases with $q_T$, approaching $90^\circ$
for $q_T >> Q$. Figure~\ref{fig2}
also shows the ``lepton plane" formed by the
momentum vector of $l^-$ and the
$\hat z$ axis. The $l^-$ and $l^+$ are emitted back-to-back
with equal momenta in the rest frame of $Z$.

Viewed from its rest frame, the $Z$ boson must be formed via
the annihilation of a pair of collinear $q$ and $\bar q$
with equal momenta, as illustrated in Fig.~\ref{fig2}.
We define the momentum unit vector of $q$ as $\hat z^\prime$, and the
``quark plane" is formed by the $\hat z^\prime$ and $\hat z$ axes.
The polar and azimuthal angles of the $\hat z^\prime$ axis
are denoted as $\theta_1$
and $\phi_1$, respectively. It is important to note that the $l^-$
angular distribution must be azimuthally symmetric
with respect to
the $\hat z^\prime$, namely,
\begin{equation}
\frac{d\sigma}{d\Omega} \propto  1 + a \cos \theta_0 + \cos^2\theta_0,
\label{eq:eq3}
\end{equation}
where $\theta_0$ is the angle between the $l^-$
momentum vector and the $\hat z^\prime$ axis (see Fig.~\ref{fig2}),
and $a$ is the
forward-backward asymmetry originating from the parity-violating
coupling to the $Z$ boson. Equation (3) shows that the lepton
angular distribution has a very simple form when measured with respect
to the $q \bar q$ axis.

As $\theta_0$ is, in general, not an experimental observable, the cross
section must be expressed in terms of the observables $\theta$ and
$\phi$. This can be accomplished by using the relation
\begin{equation}
\cos \theta_0 = \cos \theta \cos \theta_1 + \sin \theta \sin \theta_1
\cos (\phi - \phi_1).
\label{eq:eq4}
\end{equation}
Substituting Eq.~(\ref{eq:eq4}) into Eq.~(\ref{eq:eq3}), we obtain the
following expression:
\begin{eqnarray}
\frac{d\sigma}{d\Omega} & \propto & (1+\cos^2\theta)+
\frac{\sin^2\theta_1}{2} (1-3\cos^2\theta)\nonumber \\
& + & (\frac{1}{2} \sin 2\theta_1 \cos \phi_1)
\sin 2\theta \cos\phi \nonumber \\
& + & (\frac{1}{2} \sin^2\theta_1 \cos 2\phi_1)
\sin^2\theta \cos 2\phi \nonumber \\
& + & (a \sin \theta_1 \cos \phi_1) \sin\theta \cos\phi
+ (a \cos \theta_1) \cos\theta \nonumber \\
& + & (\frac{1}{2} \sin^2\theta_1 \sin 2\phi_1) \sin^2\theta \sin 2\phi
\nonumber \\
& + & (\frac{1}{2} \sin 2\theta_1 \sin\phi_1) \sin 2\theta \sin\phi
\nonumber \\
& + & (a \sin\theta_1 \sin\phi_1) \sin\theta \sin\phi,
\label{eq:eq5}
\end{eqnarray}
which is of the same form as Eq.~(\ref{eq:eq1}).
A comparison between Eq.~(\ref{eq:eq1}) and
Eq.~(\ref{eq:eq5}) shows that $A_i$ can be expressed
in terms of the three quantities, $\theta_1$, $\phi_1$ and $a$, as follows:
\begin{align}
A_0 &= \langle \sin^2\theta_1 \rangle & A_1 &= \frac{1}{2} \langle \sin
2\theta_1 \cos \phi_1 \rangle \nonumber \\
A_2 &=  \langle \sin^2\theta_1 \cos 2\phi_1 \rangle & A_3 &= \langle a
\sin \theta_1 \cos \phi_1 \rangle \nonumber \\
A_4 &=  \langle a \cos \theta_1 \rangle & A_5 &=  \frac{1}{2}
\langle \sin^2\theta_1 \sin 2\phi_1 \rangle \nonumber \\
A_6 &= \frac{1}{2} \langle \sin 2\theta_1 \sin\phi_1 \rangle &
A_7 &=  \langle a \sin\theta_1 \sin\phi_1 \rangle .
\label{eq:eq6}
\end{align}
Equation~(\ref{eq:eq6}) is a generalization of an earlier work~\cite{oleg}
which considered the special case of $\phi_1 = 0$ and
$a = 0$. The $\langle \cdot \cdot \rangle$ in Eq.~(\ref{eq:eq6})
is a reminder that the measured values of $A_i$ are averaged over the events.

As shown in Eq.~(\ref{eq:eq6}), the $q_T$ and $y$ dependencies of the
angular distribution coefficients, $A_i$, are entirely governed by the
$q_T$ and $y$ dependencies of $\theta_1, \phi_1$ and $a$.
We now consider the quantities $\theta_1$ and $\phi_1$. At the
leading-order ($\alpha_s^0$), the quark axis, $\hat z^\prime$,
is collinear with the beam axis. Hence, the result
$\theta_1 = 0$ (or $\theta_1 = \pi$) is obtained,
and Eq.~(\ref{eq:eq6})
shows that all $A_i$ except $A_4$ vanish.

\begin{figure}[tb]
\centering
\subfigure[]
{\includegraphics*[width=0.23\textwidth]{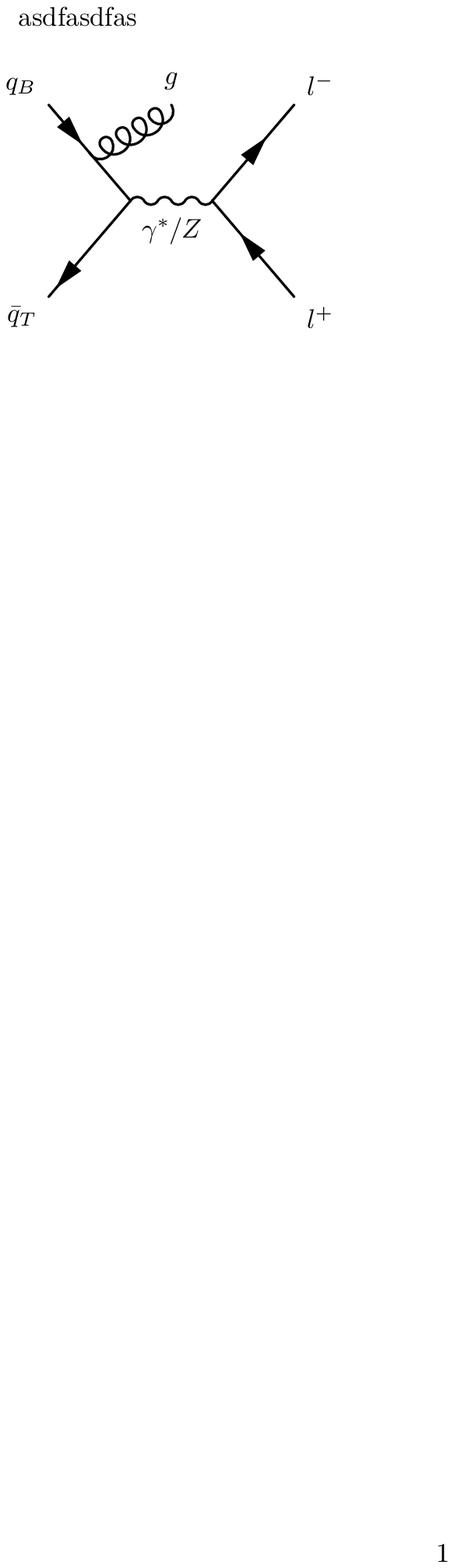}}
\subfigure[]
{\includegraphics*[width=0.23\textwidth]{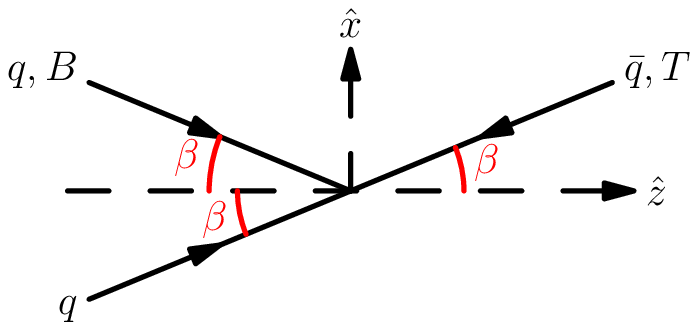}}
\subfigure[]
{\includegraphics*[width=0.23\textwidth]{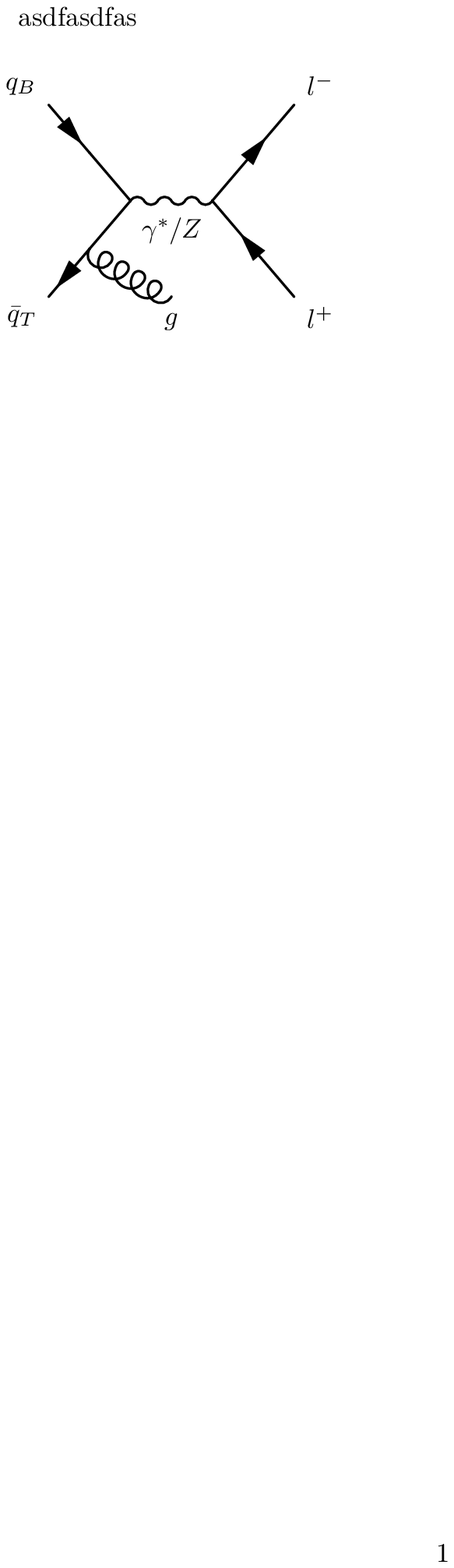}}
\subfigure[]
{\includegraphics*[width=0.23\textwidth]{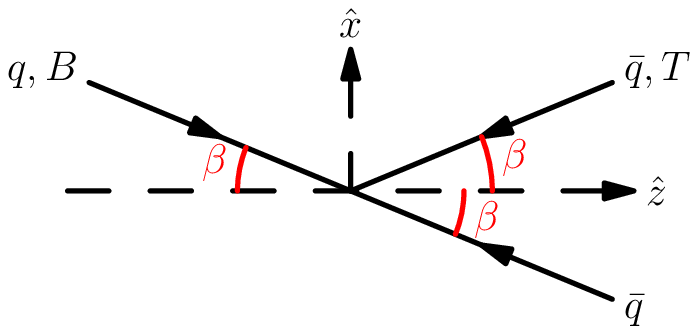}}
\caption{(a) Feynman diagram for $q \bar q$ annihilation where a gluon
is emitted from a quark in the hadron $B$. (b) Momentum direction for
$q$ and $\bar q$ in the C-S frame before and after gluon emission.
Initially, the $q$ and $\bar q$ are collinear with the hadron
$B$ and $T$, respectively. After gluon emission,
$q$ and $\bar q$ become collinear. Note that the $q$ and $\bar q$
always make an angle $\beta$ with respect to the $\hat z$ axis in
the C-S frame. (c) Feynman
diagram for the case where a gluon is emitted from an antiquark in the
hadron $T$. (d) Momentum direction for $q$ and $\bar q$ in the C-S
frame before and after gluon emission for diagram (c).
Again, $q$ and $\bar q$ become collinear after gluon emission.}
\label{fig3}
\end{figure}

At the next-to-leading order (NLO),
$\alpha_s$, a hard gluon or quark (antiquark) is emitted
so that $Z$ acquires nonzero $q_T$. Figure~\ref{fig3}(a) shows
the Feynman diagram for the $q \bar q$ annihilation process
in which a gluon is emitted from the quark in hadron $B$.
Figure~\ref{fig3}(b) shows that, initially, the $q$ and $\bar q$ are
moving collinearly with the hadron $B$ and $T$, respectively, making
an angle $\beta$ with respect to the $\hat z$ axis. After the gluon
emission, the momentum vector of the $q$
is modified such that it is now opposite to $\bar q$'s momentum
vector in the rest frame of $Z$.
Since $\bar q$ and hadron $T$ have the same momentum direction,
the $\hat z^\prime$ axis
is along the direction of $- \vec p_T$.
From Fig.~\ref{fig2}, it is evident
that $\theta_1 = \beta$ and $\phi_1 = 0$ in this case. Similarly,
for the case of Fig.~\ref{fig3}(c), where a gluon is emitted from an antiquark
in the hadron $T$, one obtains $\theta_1 = \beta$ and $\phi_1 = \pi$,
as illustrated in Fig.~\ref{fig3}(d).
Analogous results
can be found when the roles of beam and target are interchanged.
Given $\theta_1 = \beta$ (or $\theta_1 = \pi - \beta$) and
$\tan \beta = q_T/ Q$ in the Collins-Soper frame,
Eq.(\ref{eq:eq6}) gives the following
result for the
NLO $q \bar q$ annihilation processes:
\begin{equation}
A_0 = \sin^2 \theta_1 = q^2_T / (Q^2 + q^2_T).
\label{eq:eq7}
\end{equation}
Since $\phi_1 = 0$ or $\pi$, Eq.~(\ref{eq:eq6}) shows that
the Lam-Tung relation, $A_0 = A_2$, is satisfied in this case.

\begin{figure}[tb]
\centering
\subfigure[]
{\includegraphics*[width=0.23\textwidth]{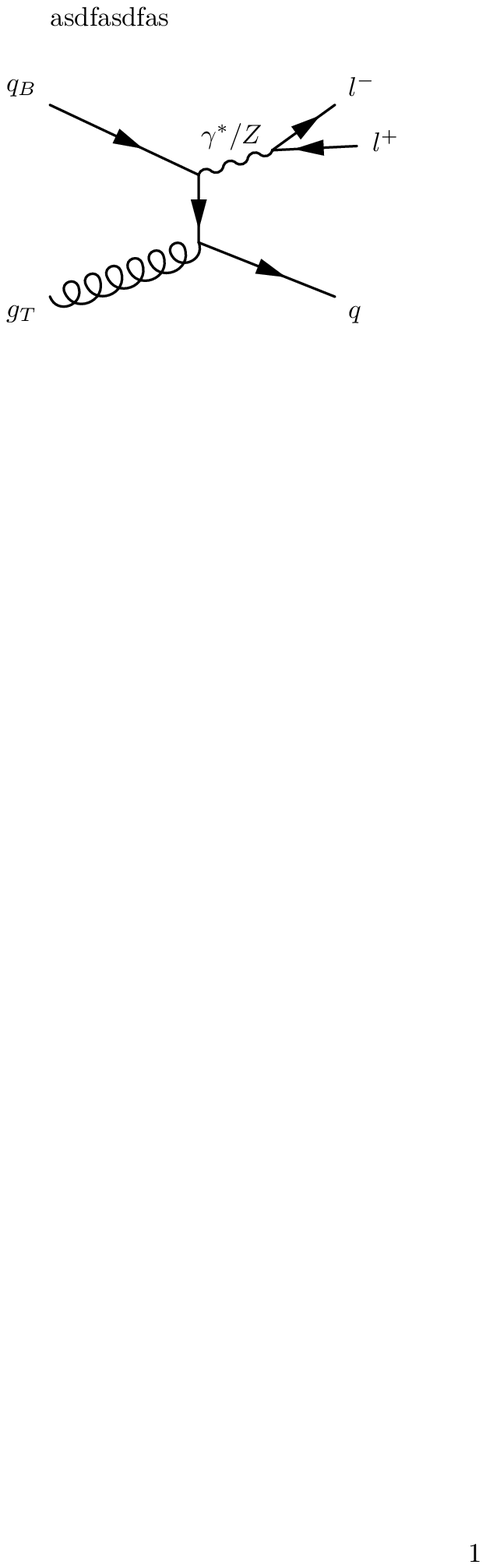}}
\subfigure[]
{\includegraphics*[width=0.23\textwidth]{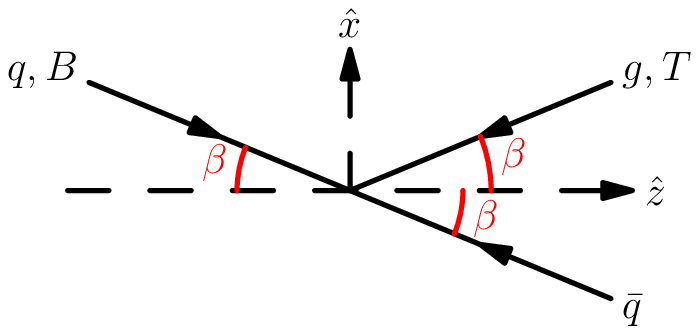}}
\subfigure[]
{\includegraphics*[width=0.23\textwidth]{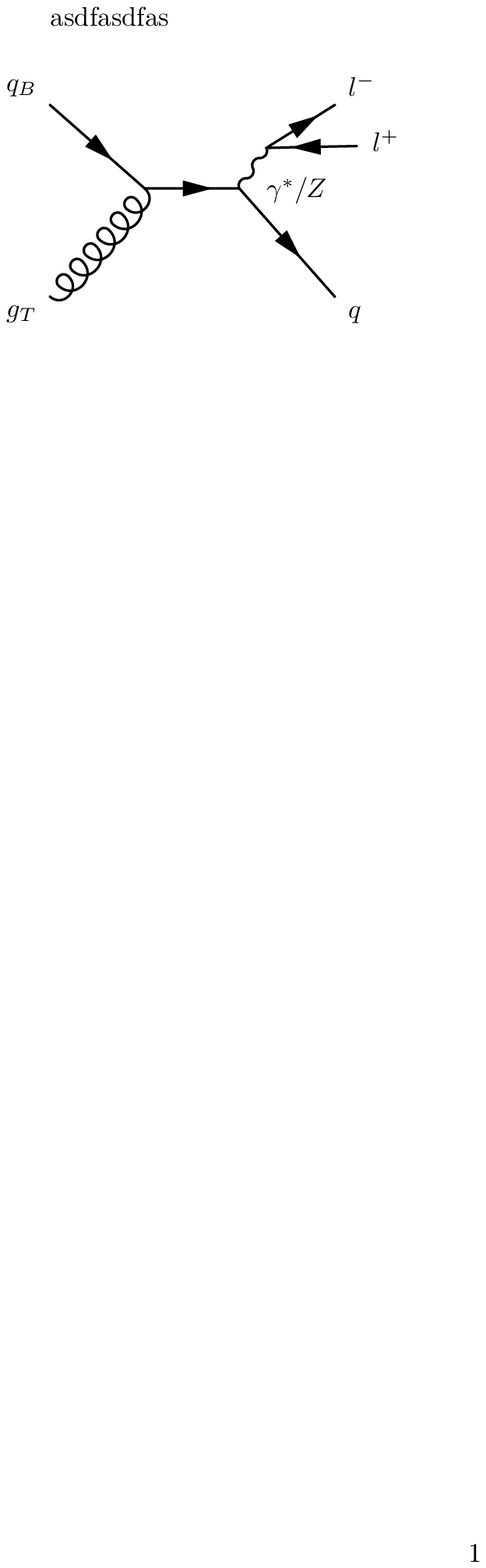}}
\subfigure[]
{\includegraphics*[width=0.23\textwidth]{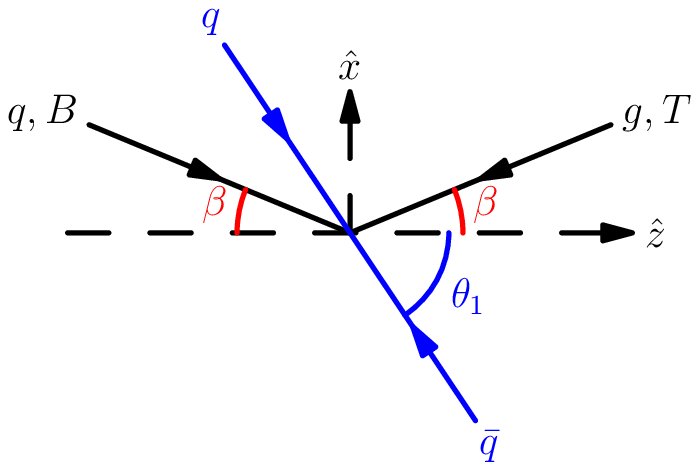}}
\caption{ (a) Feynman diagram for $qg$ Compton process where a quark from
hadron $B$ annihilates with an antiquark from the splitting of a gluon
in hadron $T$. (b) Momentum direction of $q$, $\bar q$ and $g$ in the
C-S frame before and after gluon splitting.
(c) Feynman diagram for $qg$ fusing into a quark which then emits
a $Z$. (d) Momentum direction of
$q$, $\bar q$ and $g$ before and after the $qg$ fusion.}
\label{fig4}
\end{figure}

We next consider the Compton process at NLO.
Unlike the cases for the $q \bar q$ initial state shown in
Fig.~\ref{fig3} where a hard gluon is emitted, a hard quark or
antiquark will now accompany the $Z$ in the final state.
Fig.~\ref{fig4}(a) shows the diagram in which a gluon from hadron $T$
splits into a $q \bar q$ pair and the quark from hadron $B$ annihilates
with the antiquark into a $Z$ boson. Since the momentum vector of the
quark in hadron $B$ is unchanged, $\theta_1 = \beta$ and
$\phi_1 = \pi$, as shown in Fig.~\ref{fig4}(b). This result is
identical to that
for the $q \bar q$ initial state shown in Fig.~\ref{fig3}(d).
Analogous results with $\theta_1 = \beta$ and $\phi_1 = 0$
are obtained when gluon is emitted from the
beam hadron, or when an antiquark replaces the quark in the
initial state. However, a
different situation arises, as shown in Fig.~\ref{fig4}(c),
where the quark and gluon
fuse into a quark, which then emits a $Z$. As indicated in
Fig. 4(d), $\theta_1$ must satisfy $\beta \le \theta_1 \le \pi - \beta$,
since the momenta of the initial quark and gluon combine vectorially,
resulting in a $\theta_1$ within these two limits. Therefore, the
Compton processes would lead to a $\theta_1$ larger than
$\beta$, with the exact value governed by the relative weight of these
two processes. It was shown by Thews~\cite{thews} that, to a very good
approximation, $A_0$ for the $qg$ Compton processes at order $\alpha_s$
can be given as
\begin{equation}
A_0  = 5 q_T^2 / (Q^2 + 5 q^2_T).
\label{eq:eq8}
\end{equation}
Since $\phi_1 = 0$ or $\pi$, the Lam-Tung relation, $A_0 = A_2$, is again
satisfied for the Compton process at NLO.

The dotted and dashed curves in Fig. 1(a) correspond to calculations
using Eqs. (7) and~(\ref{eq:eq8})
for the $q \bar q$ annihilation and the $qg$ Compton processes, respectively.
As the $q \bar q$ and $qg$ processes contribute to
the $p p \to Z X$ reaction incoherently, the observed $q_T$
dependence of $A_0$ reflects the combined effect of these two
contributions. A best-fit to the CMS $A_0$ data gives
a mixture of 58.5$\pm$1.6\% $qg$ and 41.5$\pm$1.6\% $q \bar q$
processes. The solid curve in Fig.~\ref{fig1}(a) shows that the data at
both rapidity regions can be well described by this mixture of the $qg$
and $q \bar q$ processes. For $pp$ collisions at the LHC,
the $qg$ process is
expected to be more important than the $q \bar q$ process, in agreement
with the best-fit result.
While the amount of $qg$ and $q \bar q$
mixture can in principle depend on the rapidity, $y$, the CMS data
indicate a very weak, if any, $y$ dependence.
The good description of $A_0$ shown in Fig.~\ref{fig1}(a) also
suggests that higher-order QCD processes
do not affect the values of $\theta_1$ significantly.

We next consider the CMS data on the $A_2$ coefficient. As shown in
Eq.~(\ref{eq:eq6}), $A_2$ depends not only on $\theta_1$,
but also on $\phi_1$. In leading order $\alpha_s$ where only a single
undetected parton
is present in the final state, the $\hat z^\prime$ axis must lie in
the hadron plane, implying $\phi_1 = 0$ and the Lam-Tung relation
is satisfied.
We first compare the CMS data, shown in Fig.~\ref{fig1}(b), with
the calculation for $A_0 = A_2$.
The dashed curve
uses the same mixture of 58.5\% $qg$ and 41.5\% $q \bar q$
components as obtained from the $A_0$ data.
The $A_2$ data are at a variance with this calculation, suggesting
the presence of higher-order QCD processes leading to a non-zero
value of $\phi_1$ (see Eq. (6)). We then performed a fit to the $A_2$
data allowing $A_2/A_0$ to be different from 1, caused by a non-zero
value of $\phi_1$. The best-fit value is $A_2/A_0 = 0.77 \pm 0.02$. The
solid curve in Fig.~\ref{fig1}(b) corresponds to the best fit to the data.
The non-zero value of
$\phi_1$ implies that the Lam-Tung relation, $A_0 = A_2$, is violated.
This violation is shown
explicitly in Fig.~\ref{fig1}(c). The solid curve obtained
with $A_2/A_0 = 0.77$ describes the observed violation of the
Lam-Tung relation well.

The violation of the Lam-Tung relation reflects the
non-coplanarity between the quark plane and the hadron
plane (i.e., $\phi_1 \ne 0$).
This can be caused by higher-order QCD processes, where
multiple partons, in addition to the detected $Z$, are present in
the final state.

\begin{figure}[tb]
\includegraphics*[width=\linewidth]{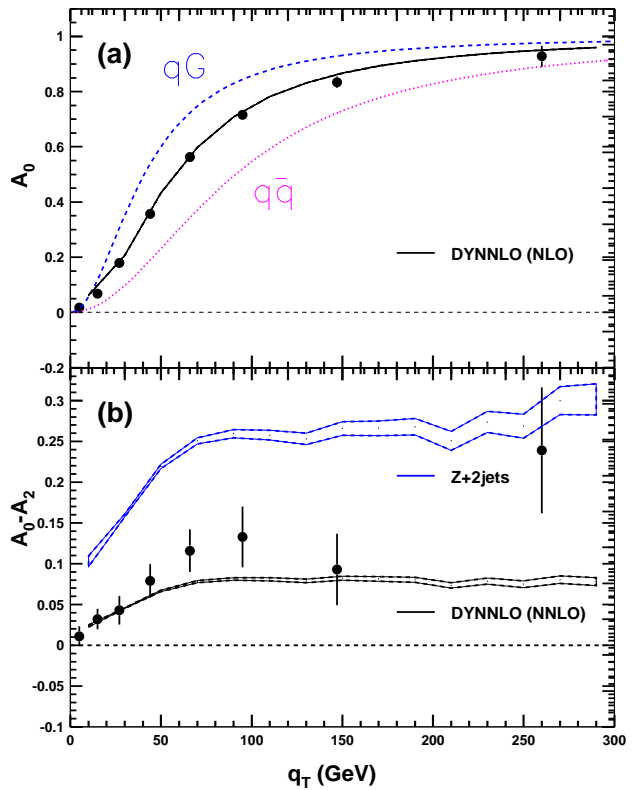}
\caption{Comparison between the CMS data on $A_0$ and
$A_0 - A_2$ with perturbative QCD calculations.
Curves correspond to calculations
described in the text.}
\label{fig5}
\end{figure}

The angular distribution results reported by the CMS Collaboration
correspond to inclusive $Z$ boson production. Based on the analysis
presented above, we expect that interesting new results would be
obtained if the data were analyzed according to the multiplicity and types
of jets accompanying the $Z$-boson. In particular, we have the following
predictions:

a) For $Z$ plus single-jet events, Fig.~\ref{fig1}(a) shows that the
$q_T$ dependence
for $A_0$ is very different between the $q \bar q$
annihilation process and the $qg$ Compton process. Since the
$q \bar q (qg)$ process contains an associated high-$p_T$ gluon (quark) jet at
the $\alpha_s$ level, as shown in Figs.~\ref{fig3} and~\ref{fig4}, one
could utilize the
existing algorithms for quark (gluon) jet identification to separate the
$q \bar q$ annihilation events from the $qg$ Compton events.
Therefore, we predict that the $Z$ plus single quark-jet events would
give a distinctly different $A_0$ from that of $Z$ plus single gluon-jet
events.
These $Z$ plus single jet $A_0$ data can also provide a powerful experimental
tool to test various algorithms for discriminating a quark jet from
a gluon jet~\cite{aad14,cms16,thaler18}.

b) As all $A_i$ coefficients depend on the values of $\theta_1$
(see Eq.~(\ref{eq:eq6})),
we expect that the $q_T$ dependence of all $A_i$, not just $A_0$, would be
different for the $q \bar q$ annihilation and the $qg$ Compton events.
This prediction can be readily tested from the existing $Z$ production data.
Furthermore, these $A_i$ angular coefficients would provide
additional experimental tools for testing the algorithms for
discriminating quark from gluon jets.

c) As discussed above, the Lam-Tung relation is expected to be valid
for $Z$ plus single-jet events. Hence, the angular distributions data
for these single jet events are predicted to satisfy $A_0 = A_2$ at all
values of rapidities and $q_T$. This remains to be tested with
the high statistics $Z$ production data from the LHC.

d) For the $Z$ plus multi-jet data, the Lam-Tung relation is expected
to be violated at a higher level than that of the inclusive $Z$ production
data. Removal of the $Z$ plus single-jet events, which
must satisfy the Lam-Tung relation, would enhance the violation of the
Lam-Tung relation.
Again, this can be tested with existing LHC data~\cite{aaboud17,cms18}.

To illustrate the points discussed above,
we have carried out perturbative
QCD calculations using the code DYNNLO~\cite{catani07,catani09}.
The parton distribution functions used in the NLO and NNLO calculations
are the CT14nlo and CT14nnlo sets.
Figure~\ref{fig5}(a) shows the comparison
between the CMS $A_0$ data at $|y|<1.0$ and the perturbative QCD
calculation
at the order $\alpha_s$. The large difference
in $A_0$ for the $q \bar q$ and $qg$ processes is consistent with
the results shown in Fig.~\ref{fig1}(a) obtained with the geometric model.
This lends support to the expectation that one can use the $Z$ plus single-jet
events to test the various jet identification algorithms.

Figure 5(b) compares the DYNNLO calculations with the CMS $A_0 - A_2$ data.
The black band corresponds to the NNLO calculation including
contributions from single jet and two jets. The blue band singles out
the contributions to $A_0 - A_2$ from $Z$ plus 2 jets only, showing
that the violation of the Lam-Tung relation is indeed amplified
for the multi-jet events. This can be readily tested with the data
collected at the LHC.

In summary, we have presented an intuitive interpretation for the lepton
angular distribution coefficients for $Z$ boson production in hadron
collision. We first derive the general expression (Eq. (5)) for the
lepton polar and azimuthal angular distribution in the $Z$ boson rest frame,
starting from the azimuthally symmetric lepton angular distribution
(Eq. (3)) with respect to the quark-antiquark axis. We show that the
various angular distribution coefficients are governed by three
quantities, $\theta_1, \phi_1$ and $a$ (Eq. (6)).
The $q_T$ dependence of $A_0$ is found to be very well described using the
leading-order results for $\theta_1$. It also allows a determination of
the relative fractions of these two processes. This result is noteworthy,
as it shows that a measurement of the angular distribution coefficient
$A_0$ alone could lead to important information on the dynamics of the
production mechanism, namely, the relative contribution of the
$q \bar q$ annihilation and the $q G$ Compton processes.

The CMS data clearly show that the Lam-Tung relation, $A_0 = A_2$, is
violated. The origin of this violation is attributed in our approach to
the deviation of $\phi_1$ from zero, indicating the non-coplanarity
between the hadron and quark planes. This non-coplanarity is caused by
higher-order QCD processes. We show that the amount of
non-coplanarity can be deduced from the $A_0 - A_2$ data directly.

We discuss how the
measurement of $A_0$ and $A_2$ coefficients in $Z$ plus single-jet
or multi-jet events
would provide valuable insight on the origin of the violation of
the Lam-Tung relation. We also show that the $A_0$ coefficient in
$Z$ plus single-jet events would be a powerful tool for testing various
algorithms which discriminate quark jets from gluon jets.

This work
was supported in part by the U.S. National Science Foundation and
the Ministry of Science and Technology of Taiwan. It was also
supported in part by the U.S. Department of Energy, Office of Science,
Office of Nuclear Physics
under contract DE-AC05-060R23177.

\end{document}